\documentclass{article}
\usepackage[T2A]{fontenc}
\usepackage[utf8]{inputenc}
\usepackage[english]{babel}
\usepackage[dvips]{graphicx}
\usepackage{caption}
\usepackage{float}
\graphicspath{{images/}}
\usepackage[toc,title,page]{appendix}

\usepackage[hcentering, bindingoffset = 10mm, right = 15 mm, left = 10 mm, top=20 mm, bottom = 20 mm]{geometry} 

\DeclareGraphicsExtensions{.HEIC,.pdf,.png,.jpg} 
\usepackage{mathtext,cite, tikz, amsmath, amsfonts, amssymb, amsthm, mathtools, multirow, lipsum, amsmath, amstext, subcaption, upgreek, wrapfig, adjustbox, icomma, amssymb, enumerate, indentfirst, float, mathrsfs, hyperref, gensymb, chngpage, pgfplots, hhline, tabularx, xfrac, tikz-feynman, physics}
\usepackage{physics}
\title{On the $K^+ \to \pi^0\pi^0\pi^0 e^+ \nu$ decay.}
\author{ \bfseries E. K. Karkaryan   \footnote{karkaryan@bk.ru}\\
 \textit{I.E. Tamm Department of Theoretical Physics, Lebedev Physical Institute,}\\
 \textit{53 Leninskiy Prospekt, Moscow, 119991, Russia} \\
 [2ex]
 \bfseries K. V. Kiselev\\
\textit{Moscow Institute of Physics and Technology, Dolgoprudny 141700, Russia}\\[2ex]
\bfseries V. F. Obraztsov\\
 \textit{NRC "Kurchatov Institute" -IHEP,}\\
 \textit{142281 Protvino, Russia}\\[2ex]
 \bfseries I. V. Surkov \\ 
\textit{Moscow Institute of Physics and Technology, Dolgoprudny 141700, Russia} \\[2ex]
\bfseries M. I. Vysotsky\\
\textit{I.E. Tamm Department of Theoretical Physics, Lebedev Physical Institute,}\\
 \textit{53 Leninskiy Prospekt, Moscow, 119991, Russia} \\
 [2ex]}
\date{}

\begin{document}

\maketitle

\begin{abstract}
    The OKA Collaboration has obtained a new upper limit on the $K^+ \to 
 \pi^0\pi^0\pi^0 e^+ \nu$ decay probability which is $65$ times lower than the one currently listed by PDG. However it is still about $10^4$ times worse than the theoretical prediction. It was suggested that production of pionium $A_{2\pi}$ in the final state of the semileptonic $K^+$ decay with subsequent $A_{2\pi}\to \pi^0\pi^0$ decay will increase the value of theoretical branching ratio due to the absence of five-particle phase space suppression. We demonstrate that despite the gain in the phase space the decay $K^+ \to  A_{2\pi}\pi^0 e^+ \nu$ appears to be strongly suppressed because of the smallness of the pionium wave function at zero $\pi$ mesons separation.
\end{abstract}

In the recent paper \cite{oka} the experimental bound $Br\qty(K^+ \to \pi^0\pi^0\pi^0 e^+ \nu)<5.4\times 10^{-8}$ at the 90$\%$ confidence level was obtained, that is 65 times lower than the value currently listed by \cite{rpp}. However the 
 estimate $Br\qty(K^+ \to \pi^0\pi^0\pi^0e^+\nu) = 2.5 \times 10^{-12}$ \cite{blaser} obtained within the chiral perturbation theory is too small for the experimental observation of this decay. As it is emphasized in \cite{blaser} the small value of the decay probability under consideration with respect to the probability of the experimentally observed decay $K^+ \to \pi^0\pi^0e^+\nu$ is due to the five-particle phase volume suppression.

\begin{figure}[H]
    \centering  \includegraphics[width=0.3\linewidth]{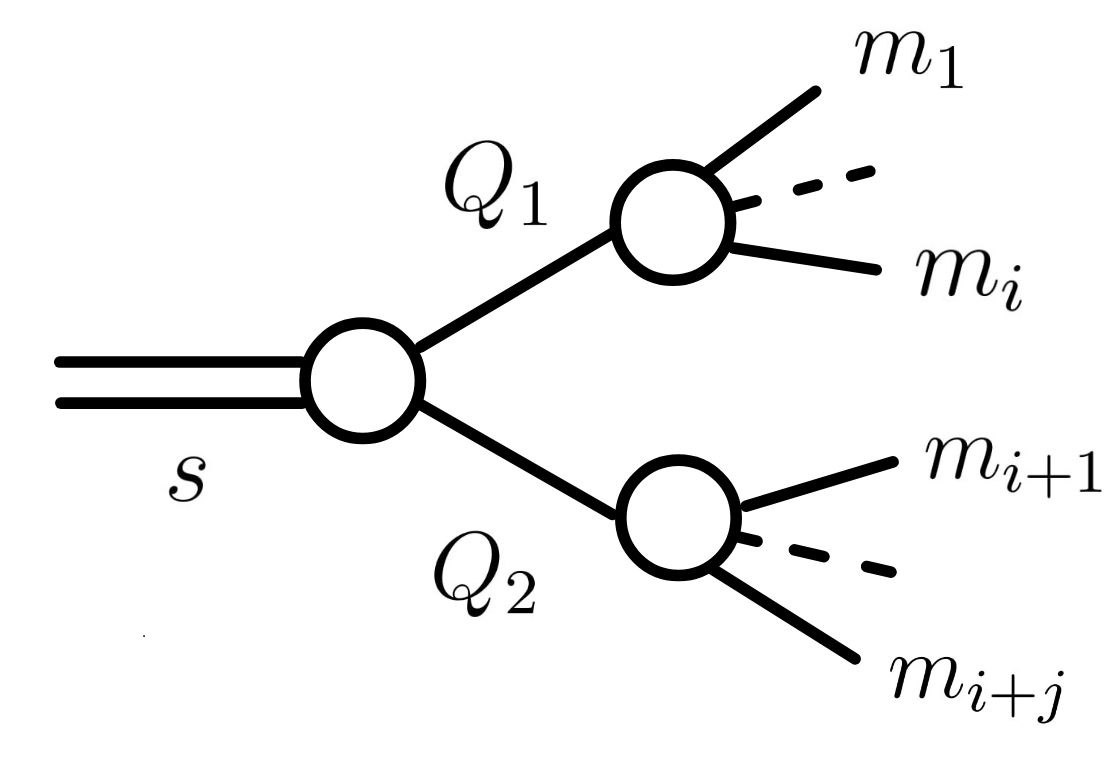}
    \caption{Illustration of the recurrent formula \eqref{1}}
    \label{fig:enter-label_1}
\end{figure}

\begin{figure}[H]
    \centering
\includegraphics[width=0.3\linewidth]{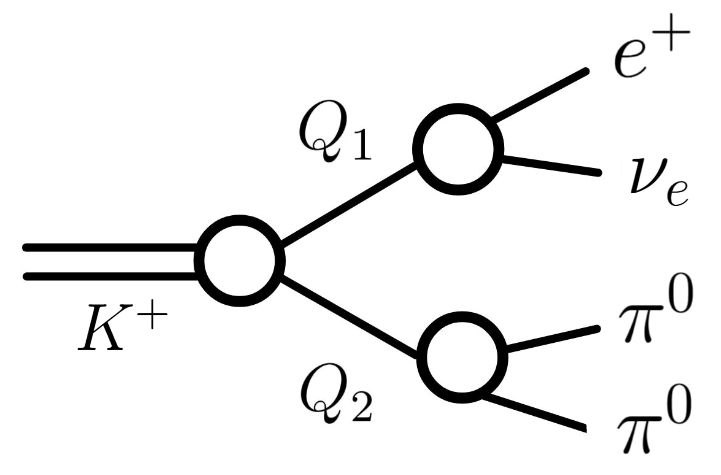}
    \caption{Illustration of the formula \eqref{2}}
    \label{fig:enter-label_2}
\end{figure}

Let us compare four- and five-particle phase volumes starting with the  calculation of the four-particle phase volume characterizing $K^+ \to \pi^0\pi^0e^+\nu$ decay. To perform the calculation the recurrent formula allowing to reduce the $n$-particle phase volume to the product of $i$- and $j$-particle ones ($n=i+j$)\cite{kayanti} is used  (see Fig.\ref{fig:enter-label_1}):
\begin{equation}\label{1}
    dV_n\qty(s; m^2_1, m^2_2,..., m^2_n) = \int \frac{dQ^2_1}{2\pi} \frac{dQ^2_2}{2\pi} dV_2(s; Q^2_1, Q^2_2)\times dV_i\qty(Q^2_1; m^2_1,..., m^2_i) \times dV_j\qty(Q^2_2; m^2_{i+1},..., m^2_{i+j}).
\end{equation}
Denoting the sum of $e^+$ and $\nu$ momenta as $Q_1$, and the sum of two $\pi^0$ mesons momenta as $Q_2$, neglecting the positron and neutrino masses and using the expression for the two-particle phase volume (see Fig.\ref{fig:enter-label_2}), we obtain: 
\begin{equation}\label{2}
    V_4(s=m^2_K; 0, 0, m^2_{\pi^0}, m^2_{\pi^0}) = \int \frac{dQ^2_1}{2\pi} \frac{dQ^2_2}{2\pi} \frac{\sqrt{\qty[s - \qty(\sqrt{Q^2_1} - \sqrt{Q^2_2})^2]\qty[s - \qty(\sqrt{Q^2_1} + \sqrt{Q^2_2})^2]}}{8\pi s} \times \frac{1}{8\pi} \times \frac{\sqrt{Q^2_2\qty(Q^2_2 - 4m^2_{\pi})}}{8\pi Q^2_2}.
\end{equation}
The integration should be performed in the following domain: $\qty(m_K - 2m_{\pi^0})^2 > Q^2_1 > 0$, $\qty(m_K - \sqrt{Q^2_1})^2 > Q^2_2 > 4m^2_{\pi^0}$.
Substituting dimensionless variables $x = Q^2_1/m^2_K$, $y = Q^2_2/m^2_K$ we get:
\begin{equation}
    V_4 = \frac{m^4_K}{2^{11} \pi^5} \int \limits^{(1 - 2m_{\pi^0}/m_K)^2}_0 dx \int\limits^{(1-\sqrt{x})^2}_{(2m_{\pi^0}/m_K)^2} dy \sqrt{\qty[1 - \qty(\sqrt{x} - \sqrt{y})^2]\qty[1 - \qty(\sqrt{x} + \sqrt{y})^2]} \sqrt{1 - \frac{\qty(4m_{\pi^0}/m_K)^2}{y}} = 7.1\times 10^{-3} \frac{m^4_K}{2^{11} \pi^5}.
\end{equation}

\begin{figure}[H]
    \centering
    \includegraphics[width=0.3\linewidth]{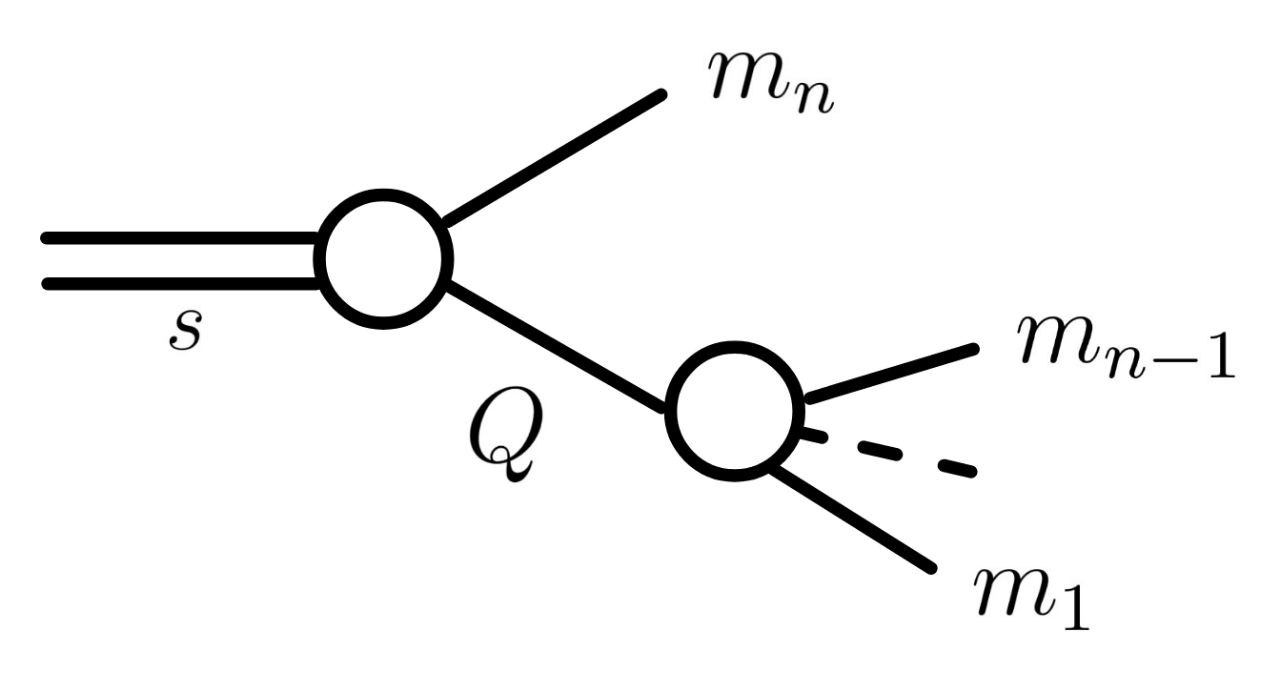}
    \caption{Illustration of the recurrent formula \eqref{4}}
    \label{fig:enter-label_3}
\end{figure}

For the calculation of the five-particle phase volume, describing the $K^+ \to \pi^0\pi^0\pi^0 e^+ \nu$ decay, we need the expression for the phase volume of three $\pi^0$ mesons. Taking into consideration the kinematics the non-relativistic approximation $\sqrt{s} = 3m_{\pi^0} + \Delta$, $\Delta \ll m_{\pi^0}$ can be used. Using the recurrent formula \cite{kayanti} (see Fig.\ref{fig:enter-label_3})
\begin{equation}\label{4}
    dV_n\qty(s; m^2_1, m^2_2,..., m^2_n) = \int \frac{dQ^2}{2\pi} dV_{n-1}\qty(Q^2; m^2_1,...,m^2_{n-1})\times dV_2\qty(s; Q^2, m^2_n)
\end{equation}
for the three-particle phase volume we obtain:
\begin{equation}
    V_3 = \int\limits^{\qty(\sqrt{s} - m_3)^2}_{\qty(m_1+m_2)^2} \frac{dQ^2}{2\pi} \frac{\sqrt{\qty[Q^2 - \qty(m_1-m_2)^2]\qty[Q^2 - \qty(m_1+m_2)^2]}}{8\pi Q^2} \frac{\sqrt{\qty[s - \qty(\sqrt{Q^2}-m_3)^2]\qty[s - \qty(\sqrt{Q^2}+m_3)^2]}}{8\pi s}.
\end{equation}
Assuming $m_1 = m_2 = m_3 = m_{\pi^0}$ we get:
\begin{equation}
    V_3 = \frac{\qty(\sqrt{s} - 3m_{\pi^0})^2}{2^6 3\sqrt{3}\pi^2}.
\end{equation}

\begin{figure}[H]
    \centering
\includegraphics[width=0.3\linewidth]{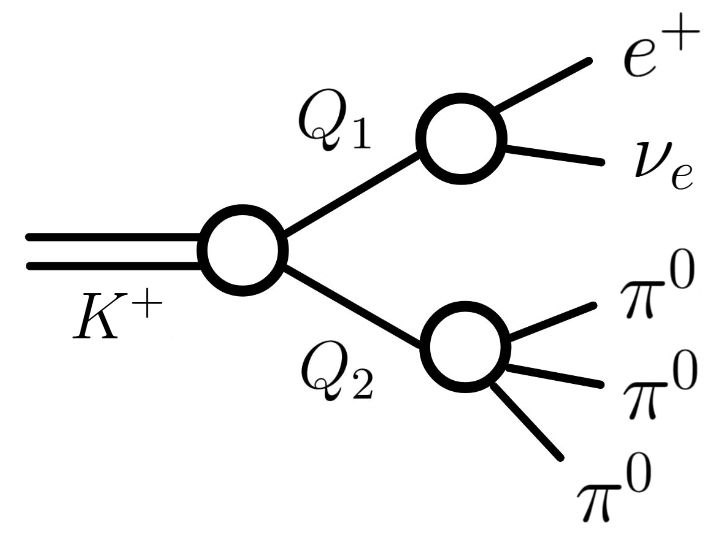}
    \caption{Illustration of the formula \eqref{7}}
    \label{fig:enter-label_4}
\end{figure}

For the five-particle phase volume using \eqref{1} we obtain (see Fig.\ref{fig:enter-label_4}):
\begin{equation}\label{7}
    V_5(m^2_K; 0, 0, m^2_{\pi^0}, m^2_{\pi^0}, m^2_{\pi^0}) = \int \frac{dQ^2_1}{2\pi} \frac{dQ^2_2}{2\pi} \frac{\sqrt{\qty[s - \qty(\sqrt{Q^2_1} - \sqrt{Q^2_2})^2]\qty[s - \qty(\sqrt{Q^2_1} + \sqrt{Q^2_2})^2]}}{8\pi s} \times \frac{1}{8\pi} \times \frac{\qty(\sqrt{Q^2_2}-3m_{\pi^0})^2}{2^6 3\sqrt{3}\pi^2}.
\end{equation}
In this case the integration should be performed in the domain $\qty(m_K - 3m_{\pi^0})^2 > Q^2_1 >0$, $\qty(m_K - \sqrt{Q^2_1})^2 > Q^2_2 > 9m^2_{\pi^0}$. In terms of dimensionless variables we obtain: 
\begin{align}
    V_5 = \frac{m^6_K}{2^{14}3\sqrt{3}\pi^6}\int\limits^{\qty(1-3m_{\pi^0}/m_K)^2}_0 dx \int\limits^{\qty(1-\sqrt{x})^2}_{\qty(3m_{\pi^0}/m_K)^2} dy& \qty(\sqrt{y} - \frac{3m_{\pi^0}}{m_K})^2 \sqrt{\qty[1 - \qty(\sqrt{x} - \sqrt{y})^2]\qty[1 - \qty(\sqrt{x} + \sqrt{y})^2]} = \\ \nonumber
    &=1.37\times 10^{-6}\frac{m^6_K}{2^{14}3\sqrt{3}\pi^6} .
\end{align}

Finally for the ratio of the phase volumes taking into account that $\pi^0$ mesons are identical we get:
\begin{equation}
    \frac{m^2_K V_4/2!}{V_5/3!} = 2.1\times 10^6.
\end{equation}
Within the chiral perturbation theory for the widths ratio it was obtained in \cite{blaser}:
\begin{equation}
    \frac{\Gamma\qty(K^+ \to \pi^0\pi^0e^+\nu)}{\Gamma\qty(K^+ \to \pi^0\pi^0\pi^0e^+\nu)} \approx 3.4\times 10^6.
\end{equation}
Thereby the relative suppression of the $K_{e5}$ decay widths originates from the smallness of the five-particle phase volume. 

Thus in the paper \cite{oka} it was suggested, that pionium $A_{2\pi}$\footnote{bound by the Coulomb interaction $\pi^+\pi^-$-atom \cite{pionium, adeva}.} production in the $K^+ \to \pi^0  A_{2\pi}e^+ \nu$ decay with the subsequent decay $A_{2\pi} \to \pi^0 \pi^0$ will lead to the considerable increase of the $K^+ \to \pi^0\pi^0\pi^0 e^+ \nu$ decay probability, since the five-particle phase volume is replaced by four-particle one in this case.

\begin{figure}
     \centering
     \includegraphics[width=0.3\linewidth]{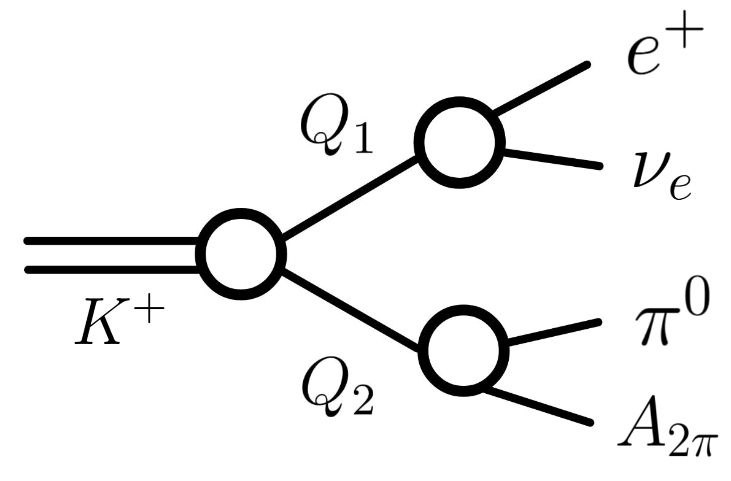}
     \caption{Illustration of the formula \eqref{11}}
     \label{fig:enter-label_5}
 \end{figure}

Let us calculate the phase volume of the $\pi^0  A_{2\pi}e^+ \nu$ state, formed in the $K^+$ meson decay. The formula \eqref{2} is replaced by the following one (see Fig.\ref{fig:enter-label_5}):
\begin{align}\label{11}
    V_A = \int \frac{dQ^2_1}{2\pi} \frac{dQ^2_2}{2\pi} \frac{\sqrt{\qty[s - \qty(\sqrt{Q^2_1} - \sqrt{Q^2_2})^2]\qty[s - \qty(\sqrt{Q^2_1} + \sqrt{Q^2_2})^2]}}{8\pi s} \times \frac{1}{8\pi}\times \\ \nonumber \times \frac{\sqrt{\qty[Q^2_2 - \qty(\sqrt{4m^2_{\pi^+}} - \sqrt{m^2_{\pi^0}})^2]\qty[Q^2_2 - \qty(\sqrt{4m^2_{\pi^+}} + \sqrt{m^2_{\pi^0}})^2]}}{8\pi Q^2_2},
\end{align}
where we took into account that the pionium mass is close to the sum of the two charged pions masses. In the dimensionless variables we obtain:
 \begin{align}\label{12}
     V_A =\frac{m^4_K}{2^{11}\pi^5}&\int\limits^{\qty(1-\qty(2m_{\pi^+}+m_{\pi^0})/m_K)^2}_0dx \int\limits^{\qty(1-\sqrt{x})^2}_{\qty(\qty(2m_{\pi^+}+m_{\pi^0})/m_K)^2} dy \sqrt{\qty[1- \qty(\sqrt{x}-\sqrt{y})^2]\qty[1- \qty(\sqrt{x}+\sqrt{y})^2]}\times \\ \nonumber\times &\sqrt{\qty[1-\frac{\qty(2m_{\pi^+} - m_{\pi^0})^2}{ym^2_K}]\qty[1-\frac{\qty(2m_{\pi^+} + m_{\pi^0})^2}{ym^2_K}]} = 0.83\times 10^{-4}\frac{m^4_K}{2^{11}\pi^5}.
 \end{align}
It leads to the increase of the decay probability by a factor 
 \begin{equation}
     \frac{m^2_K V_A}{V_5/3!} \approx5\times 10^{4},
 \end{equation}
 which is in agreement with the hypothesis of the work  \cite{oka}. Multiplying the theoretical result by this factor we get the probability of the decay exceeding about two times the upper limit obtained in the experiment \cite{oka}(!). However the amplitude of the pionium production is proportional to its wave function at zero $\pi$ mesons separation $\psi(0)=1/\sqrt{\pi a^3_B} \sim \alpha^{3/2}/\qty(2\sqrt{2\pi})$. It leads to the multiplication of this decay probability by a factor $\alpha^3/\qty(8\pi) \approx 1.5\times 10^{-8}$. In this way for the pionium contribution into the branching of $K^+ \to \pi^0\pi^0\pi^0e^+\nu$ we finally obtain:

 \begin{equation}
     \Delta Br(K^+ \to \pi^0\pi^0\pi^0e^+\nu) = \qty(2.5\times 10^{-12})  \cdot \qty(5\times10^{4}) \cdot \qty(1.5\times 10^{-8}) \approx 2 \times 10^{-15},
 \end{equation}
which is three orders of magnitude less than estimated in \cite{blaser} branching ratio of $K^+ \to \pi^0\pi^0\pi^0e^+\nu$ decay. Thus pionium contribution is completely negligible.

\end{document}